\documentclass[a4paper,11pt]{article}
\usepackage{pos}
\usepackage[utf8]{inputenc}

\usepackage{amsmath}
\usepackage{color}
\usepackage{comment}
\usepackage{graphicx}
\usepackage{epstopdf}
\usepackage{tikz}
\usepackage{graphicx}

\usepackage[english]{babel}

\DeclareMathOperator{\arcch}{arccosh}

\makeatletter

\newcommand{\be}{\begin{equation}}
\newcommand{\ee}{\end{equation}}
\newcommand{\beq}{\begin{eqnarray}}
\newcommand{\eeq}{\end{eqnarray}}
\newcommand{\ba}{\begin{align}}
\newcommand{\ea}{\end{align}}

\newcommand{\argo}{\left(\beta\momenta\right)}
\newcommand{\pressure}{\mathcal{F}}
\newcommand{\energy}{\mathcal{E}}
\newcommand{\momenta}{k_n}
\newcommand{\momentatwo}[1]{k_{#1}}
\newcommand{\mathi}{{\rm i}}

\newcommand{\nada}[1]{#1}

\title{The beauty of curved momentum space}
\ShortTitle{The beauty of curved momentum space}

\author*[a]{S.~A.~Franchino-Viñas}
\author[b,c]{S.~Mignemi}
\author[d,e]{J.~J.~Relancio}

\affiliation[a]{Helmholtz-Zentrum Dresden-Rossendorf,\\
Bautzner Landstraße 400, 01328 Dresden, Germany.}

\affiliation[b]{Dipartimento di Matematica e Informatica, Università di Cagliari,
\\
viale Merello 92, 09123 Cagliari, Italy.}

\affiliation[c]{
INFN, Sezione di Cagliari,
\\
Cittadella Universitaria, 09042 Monserrato, Italy.}

\affiliation[d]{Departamento de Matemáticas y Computación,\\
Universidad de Burgos, 09001 Burgos, Spain.}
\affiliation[e]{Centro de Astropartículas y F\'{\i}sica de Altas Energ\'{\i}as (CAPA),\\
Universidad de Zaragoza, Zaragoza 50009, Spain}

\emailAdd{s.franchino-vinas@hzdr.de}
\emailAdd{smignemi@unica.it}
\emailAdd{jjrelancio@ubu.es}

\abstract{
In this manuscript, we will discuss the notion of curved momentum space, 
as it arises in the discussion of noncommutative or doubly special relativity theories.
We will illustrate it with two simple examples, the Casimir effect in anti-Snyder space 
and the introduction of fermions in doubly special relativity.
We will point out the existence of intriguing results, 
which suggest nontrivial connections with spectral geometry and Hopf algebras.}

\FullConference{%
  Corfu Summer Institute 2022 "School and Workshops on Elementary Particle Physics and Gravity",\\
  28 August - 1 October, 2022\\
  Corfu, Greece}

\begin{document}
\maketitle


\section{Introduction}
The first notions of curved momentum space  are probably due to Snyder~\cite{Snyder:1946qz,Snyder:1947nq}.
In the quest for a Quantum Field Theory (QFT) free of infinities, 
he proposed a noncommutativity among coordinates, proportional to the Lorentz generators $\hat J_{ij}$:
\begin{align}
 [\hat x_i,\hat x_j]=\mathi\beta^2\hat J_{ij}.
\end{align}
His expectation was that a generalized uncertainty principle would then translate into 
a granular structure of spacetime, 
which, in turn, would lead to the avoidance of the ultraviolet divergences generated by the continuum structure. 
The fact is that the construction of these commutation relations was based on differential operators 
acting on an auxiliary momentum space, which can be recognized to be de Sitter space  in projective coordinates~\cite{Snyder:1946qz}.

After their  publication, these ideas were soon eclipsed by the success of the renormalization process developed in parallel by Schwinger, Feynman, and Tomonaga, among others. 
However, this success didn't prevent the sporadic appearance of related works, among them the prominent discussions by Gol'fand in the 60's,
where a quantum field theory was discussed departing from an action written in a maximally symmetric curved momentum space~\cite{Golfand:1959vqx, Golfand:1962kjf}.
Further refinements were later provided by the Russian school lead by Mir-Kasimov
\cite{Mir-Kasimov:1966a, Mir-Kasimov:1966b, Mir-Kasimov:1967, Kadyshevsky:1983yc,Donkov:1984fj}.
  
It was not until the 90's that noncommutativity, based on several grounds, 
began to gain a preponderant place in the community of high energy physics.
On one side, we can mention the idea of Connes and collaborators to obtain the standard model from a spectral action of a noncommutative space~\cite{Connes:1990qp,Chamseddine:2022rnn}.
On the other side, almost contemporarily, the study of quantum groups  provided a deformation of the Poincaré algebra~\cite{Lukierski:1991pn};
the latter became famous as $\kappa$-Poincaré and motivated the surge of Doubly Special Relativity (DSR)~\cite{AmelinoCamelia:2008qg},
a generalized version of Special Relativity (SR) with two invariant quantities.
Still a third pillar was constituted  by the discovery in string theory of structures involving a certain noncommutativity~\cite{Seiberg:1999vs, Douglas:2001ba}.

During this noncommutative eruption, the notion of momentum space was revalued~\cite{Amelino-Camelia:1999jfz, Majid:1999tc},
in particular in connection with $\kappa$-Minkowski space~\cite{KowalskiGlikman:2002ft}.
It seems, however, that the use of a curved momentum space has been by far more developed in physical 
applications rather than in formal considerations.

In the following we will try to stimulate the latter by discussing two simple physical systems: 
the Casimir energy of a scalar field in anti-Snyder space and the introduction of fermions in curved momentum space. 
They will provide hints that some mathematical notions behind curved momentum spaces may have passed unnoticed in the literature
and may thus still be there waiting for us.


\section{Casimir energy in Snyder space}
Let us begin by discussing the Casimir energy  in Snyder space.
The Casimir effect corresponds to the class of quantum vacuum effects in external backgrounds.
It was predicted by Casimir a long time ago~\cite{Casimir:1948dh};  
he realized that, due to quantum fluctuations, a force will be generated between two parallel metallic plates 
when placed in the otherwise empty spacetime.
Much attention has been devoted to it during the last decades, 
mainly after its experimental confirmation by Lamoreaux in 1997~\cite{Lamoreaux:1996wh}.
Given the vast bibliography on the subject, we limit ourselves to mentioning the reviews~\cite{Bordag:2009zz, Lamoreaux:2005gf, Milton:2001yy, Mostepanenko:1990ceg}, 
where several references are gathered, 
as well as some of its late applications to constrain dark matter and alternative theories of gravity~\cite{Klimchitskaya:2021lak,Fichet:2017bng}.

One of the most physical ways to compute the Casimir force (and actually the one employed by Casimir in his original paper) 
is to sum up all the zero-point energies of the field's oscillation modes, 
which gives the total vacuum energy as a function of some geometric quantities.
Remarkably, this implies that it possesses a close connection with spectral functions~\cite{Elizalde:1995hck, Kirsten:2001wz}.
As several quantities in QFT, 
the sum of over the frequencies turns out to be infinite, given that the frequency of the oscillators grows indefinitely; 
hence a regularization and a subsequent renormalization is usually required. 

At this point, we recall the primordial idea by Snyder~\cite{Snyder:1946qz,Snyder:1947nq},
who in an attempt to avoid such divergences introduced the notion of noncommutative spacetime:
coordinates might no longer commute, hopefully giving rise to some natural ultraviolet cutoff.
Having this in mind, the computation of the Casimir energy in noncommutative spaces turns out 
to be a test of fundamental principles. 

Some previous studies of the vacuum energies include the consideration of Moyal flat spaces~\cite{Casadio:2007ec, Fosco:2007tn}, Moyal curved spaces~\cite{Chaichian:2001pw}, $\kappa$-Minkowski space~\cite{Harikumar:2019hzq} (computing the force directly from the energy-momentum tensor in a perturbative approach on the noncommutative parameter) and the whole Snyder space~\cite{Mignemi:2017yhd}.
In the following we will review a possible way to compute the Casimir energy for a scalar field in an anti-Snyder space, first developed in~\cite{Franchino-Vinas:2020umq};
the natural language will turn out to be that of curved momentum space. 


\subsection{Euclidean Anti-Snyder space}

Recall that the Euclidean anti-Snyder space is a noncommutative space in which the algebra of position ($x_i$) and momentum ($p_j$) operators is the following:
\begin{eqnarray}\label{Snydercomm}
&&[\hat J_{ij},\hat J_{kl}]=\mathi\left(\delta_{ik}\hat J_{jl}-\delta_{il}\hat J_{jk}-\delta_{jk}\hat J_{il}+\delta_{lj}\hat J_{ik}\right),\cr
&&[\hat J_{ij},\hat p_k]=\mathi\left(\delta_{ik}\hat p_j-\delta_{jk}\hat p_i\right),\qquad[\hat J_{ij},\hat x_k]=\mathi\left(\delta_{ik}\hat x_j-\delta_{jk}\hat x_i\right),\cr
&&[\hat x_i,\hat p_j]=\mathi\left(\delta_{ij}-\beta^2\hat p_i\hat p_j\right),\qquad[\hat x_i,\hat x_j]=-\mathi\beta^2\hat J_{ij},\qquad[\hat p_i,\hat p_j]=0.
\end{eqnarray}
In these expressions, the indices $i$ and $j$ run from 1 to $D$, where $D$ is the number of dimensions
in which we are working; 
additionally, the operators $\hat J_{ij}$ generate the $\mathfrak{so}(D)$ algebra (the Lorentz algebra in a space with Euclidean signature) 
and retain their usual form when written in terms of the position and momentum operators~\cite{Mignemi:2011gr}, 
i.e. $\hat J_{ij}=\hat x_i\hat p_j-\hat x_j\hat p_i$.
Notice that the noncommutativity is enclosed in the parameter $\beta$, 
which is heuristically expected to be of the order of Planck's mass.
The fact that we are working with an anti-Snyder space is determined by $\beta$: 
if one analytically continues $\beta\to\mathi \beta$, then one obtains the algebra of Snyder space; 
below we will see how this difference manifests itself in the computations.

Instead of working at an algebraic level, we will solve the quantum equations of motion of the field employing realizations of the anti-Snyder algebra. 
Since the momentum operators commute, a simple Hilbert space 
for the realizations is that of $L_2(\mathbb{C})$ functions in momentum space, where the coordinates are denoted by $p_i$ (without a hat)
and the position operators will act as differential operators.
The realizations that we are going to employ are:
\begin{enumerate}
 \item the \emph{Snyder realization}~\cite{Snyder:1946qz,Lu:2011fh}, where we have\footnote{For the sake of clarity, at this point we will (arbitrarily) employ subindices and superindices to distinguish the coordinates in both representations. Later we will drop them to simplify the notation. } 
 \begin{align}\label{eq:realization2}
 \hat{p}_i\equiv p_i^{(S)},\quad \hat{x}_i\equiv \mathi\left(\delta_{ij}-\beta^2p_i^{(S)}p_j^{(S)}\right)  \frac{\partial}{\partial p_j^{(S)}};
\end{align}
the measure in this space, if one desires these operators to be symmetric~\cite{Lu:2011fh},  is then fixed to be $d\mu_{S}=\frac{d^Dp^{(S)}}{(1-\beta^2p_{(S)}^2)^{(D+1)/2}}$, 
 (we use the Einstein notation for the sum and set the limit $p_{(S)}^2:=p^{(S)}_ip_{(S)}^i< \beta^{-2}$ to avoid singularities);

\item the \emph{projective realization}~\cite{Mignemi:2011gr}, in which the operators read
\begin{align}\label{eq:realization1}
 \hat{p}_i\equiv \frac{p_i^{(P)}}{\sqrt{1+\beta^2{p^2_{(P)}}}},\qquad \hat{x}_i\equiv \mathi \sqrt{1+\beta^2p_{(P)}^2} \frac{\partial}{\partial p^{(P)}_i},
\end{align}
and the measure that guarantees the symmetry of the operators is $d\mu_{P}=\frac{d^Dp^{(P)}}{\sqrt{1+\beta^2p_{(P)}^2}}$ (now the coordinates $p^{(P)}_i$ are unbounded).

\end{enumerate}

 From these representations one realizes two things. 
The first is that the underlying momentum spaces correspond in both cases to hyperbolic spaces $\mathbb{H}_{D}$, the Euclidean version of anti-de Sitter spaces (AdS), 
and are therefore curved.
Indeed, one can \emph{a posteriori} see that the coordinates in the Snyder realization correspond to the metric
\begin{align}
 g^{(S)}_{\mu\nu}= \frac{\delta_{\mu\nu}}{1-\beta^2p_{(S)}^2}+\frac{\beta^2 p^{(S)}_\mu p^{(S)}_\nu}{\left(1-\beta^2p_{(S)}^2\right)^2},
\end{align}
while for the projective realization one has
\begin{align}
 g^{(P)}_{ij}=\delta_{ij}-\beta^2 \frac{p_i^{(P)}p_j^{(P)}}{1+\beta^2 p_{(P)}^2}.
\end{align}
In fact, the two representations are related by a unitary transformation mapping the corresponding Hilbert spaces, which is basically given by the change of coordinates
\begin{align}
    p_i^{(P)} = \frac{p_i^{(S)}}{\sqrt{1-\beta^2 p^2_{(S)}}}.
\end{align}

The second  is that even though the coordinates $p_i$ may be unbounded, the momentum operators are always bounded;
this is one of the main differences with the Snyder case. 
As we will see below, this will have some consequences for the computation of the Casimir energy.


\subsection{Confining a particle in anti-Snyder space}\label{sec:spectrum}
The very question on how to define boundaries in a noncommutative space is a subtle one~\cite{Lizzi:2003ru, Lizzi:2014pwa}, 
given that localization is precluded by the intuitive picture of a granular structure at small scales. 
However, one can impose boundary conditions by using the physical idea of confining a particle with an appropriate potential~\cite{Scholtz:2007ig, Falomir:2013vaa}.

To simplify the notation, let us consider plates situated at $x_{\perp}=\pm L$; 
this will enable us to split the space into the coordinate $x_{\perp}$, which is perpendicular to the plates, and the coordinates $x_{\parallel}$, 
which are parallel  to them.
If these plates are modeled as scalar potentials, then we will need to find the eigenfunctions of the Hamiltonian
\begin{align}\label{eq:hamiltonian}
\hat{ \mathcal{H}}_V:= \hat{p}^2+ V H(\hat{x}_{\perp}-L)+ V H(-\hat{x}_{\perp}-L), \quad V\in \mathbb{R},
\end{align}
where $H(\cdot)$ the Heaviside function and we are interested in the limit of infinite and positive $V$.

The projective realization is particularly well-suited to perform this computation for the following reason:
there exist eigenfunctions of the position operators\footnote{Contrary to the case in~\cite{Kempf:1994su}, these states are physical in the system that we analyze.}
\begin{align}
 \psi_{x_i}(p)=e^{-\mathi \frac{x_i}{\beta} \text{arcth}\left(\frac{\beta p_i}{\sqrt{1+\beta^2p^2}}\right)},\quad  \hat{x}_i\psi_{x_i}(p)=x_i \psi_{x_i}(p),
 \end{align}
 which, as we will see, can be used to build a basis of the Hilbert space.
 Since the position operators do not commute with each other, 
 one certainly cannot build a basis only from  their eigenvectors. 
 The solution is instead to consider eigenfunctions of only one coordinate (the $x_\perp$ in our case),
 with fixed momentum coordinates in the parallel directions:
\begin{align}
 \psi_{x_{\perp},q_{\parallel}}(p):= \frac{1}{\sqrt{2\pi}}\psi_{x_{\perp}}(p) \delta(p_{\parallel}-q_{\parallel}), \quad x_{\perp}\in\mathbb{R} \wedge  q_{\parallel}\in\mathbb{R}^{D-1}.
\end{align}
The states thus formed are labelled by $D$ real numbers, $x_{\perp}$ and $q_{\parallel}$; 
moreover, they are orthonormal in the sense of a continuum basis in momentum space, i.e.
 \begin{align}
  \begin{split}
\left( \psi_{x_{\perp},q_{\parallel}} ,\psi_{y_{\perp},k_{\parallel}}\right)
  :&= \int \frac{d^D p}{\sqrt{1+\beta^2p^2}} \psi^*_{x_{\perp},q_{\parallel}}(p) \psi_{y_{\perp},k_{\parallel}}(p)
  \\
  &=\delta(k_{\parallel}-q_{\parallel}) \delta(x_{\perp}-y_{\perp}).
  \end{split}
 \end{align}

 For the sake of rigor, notice also that there exists a basis of eigenvectors of the momentum operators, 
\begin{align}
 \phi_{q}(p)=\sqrt{1+\beta^2q^2}\delta (p-q), \quad q\in\mathbb{R}^D,
\end{align}
that satisfy
\begin{align}
 \hat p_i \phi_{q}(p) &=\frac{q_i}{\sqrt{1+\beta^2q^2}} \phi_{q}(p),
\end{align}
as well as a completeness relation with the covariant delta function in curved space
\begin{align}
\int \frac{d^D q}{\sqrt{1+\beta^2q^2}} \phi_q(p) \phi_q^*(p')&=\sqrt{1+\beta^2p^2}\delta^D(p-p').
\end{align}
In the absence of the potentials, they would constitute the set of eigenfunctions of the Hamiltonian. 
Once we turn on the potentials, we  have to build combinations from them that,
when projected onto $\psi_{\pm L,q_{\parallel}}$, must satisfy a certain continuity condition;
in the infinite limit, this condition simplifies to the vanishing of such projection.
To build the solutions we use the \emph{ansatz} of a linear combination of two eigenfunctions $\phi_q$, 
with opposite momenta in the perpendicular component:
\begin{align}
\left(\psi_{\pm L,q_{\parallel}} ,\; A_{q}\phi_q +B_{q} \phi_{-q_{\perp},q_{\parallel}} \right)=0;
\end{align}
the compatibility of the system then requires  a quantization condition on the labels of the eigenfunctions,
\begin{align}
  \sin\left( \frac{2L}{\beta} \text{arcth}\left( \frac{\beta q_{\perp}}{\sqrt{1+\beta^2q^2}}\right) \right)=0,
\end{align}
which can be easily solved for the perpendicular component,
\begin{align}\label{eq:spectrum}
 \beta^2 q_{\perp,n}^2 &=\sinh^2\left(\momenta \beta\right) \left(1+\beta^2q_{\parallel}^2\right),\quad n\in\mathbb{N}^{+},
\end{align}
where we have introduced the usual commutative quantized momenta $\frac{n\pi}{2L}=:\momenta$. 
As a fast check, one can readily verify that the commutative limit is the right one.


\subsection{The Casimir energy}
If one desires to compute the Casimir energy for a scalar field $\phi$ of mass $m$, 
then one should consider the generalization of the Klein--Gordon (KG) equation to the noncommutative case.
One subtlety is the fact that, once the time coordinate becomes noncommuting, unitarity issues may arise.
To avoid such problems, we will simply consider a spacetime which may be factorized into $\mathbb{R}\times \text{anti-Snyder}_D$.
Then, the corresponding KG equation is given by
\begin{align}
 (\partial_t^2+\hat{ \mathcal{H}} +m^2) \phi=0.
\end{align}
Fourier transforming the time to the frequency domain, we get the dispersion relation
\begin{align}\label{eq:dispersion2}
 \begin{split}
\omega_{q_{\parallel},n}^2&= \frac{q_{\parallel}^2+q_{\perp,n}^2}{1+\beta^2 \left(q_{\parallel}^2+q_{\perp,n}^2\right)} +m^2,
 \end{split}
\end{align}
where the parallel components are arbitrary, $q_{\parallel}\in \mathbb{R}^{D-1}$, and the perpendicular momentum $q_{\perp,n}$ are those quantized in Eq.~\eqref{eq:spectrum}.

Hitherto, we have computed the energy of every single field mode. 
As a consequence of the boundedness of momenta in anti-Snyder space, 
the energies of these modes are also bounded; 
this is contrary to what happens in the commutative case, 
where the energy of a single mode can be arbitrarily large.
The problem arises when trying to compute the sum of all these energies, 
which, in analogy to the commutative case, is expected to give 
the energy density $\energy$ per unit area in the parallel directions
($\Omega_D$ is the hypersurface of a $D$-dimensional hypersphere):
\begin{align}\label{eq:casimir}
 \energy
 &= \frac{\Omega_{D-2}}{2} \sum_{n=1}^{\infty~}\int_0^{\infty} \frac{dq}{(2\pi)^{D-1}}\,q^{D-2} \sqrt{\frac{q^2}{1+\beta^2q^2}+\frac{\tanh^2\left(\beta\momenta\right)}{\beta^2(1+\beta^2q^2)}+m^2}.
\end{align}
Indeed, one immediately notices from this expression that the divergence is due not to the contribution of high energy modes;
instead, it is a consequence of the  integration domain being unbounded. 
One could argue that for compact noncommutative manifolds, such as the fuzzy disc or fuzzy sphere~\cite{Madore:1991bw, Lizzi:2003ru, Falomir:2013vaa, Franchino-Vinas:2018gbv},
the generalized uncertainty principle induced by the noncommuting coordinates forces the number of available states to be finite.
In our present case, however, the underlying classical manifold is noncompact;
therefore, even if there should exist a minimal area, the noncompactness of the manifold 
creates a bypass and allows the presence of an infinite number of states.

These features invalidate some of the regularization methods most frequently used in the commutative case.
Consider for example a zeta function regularization~\cite{Elizalde:2007du},
in which one promotes the power of the energies in the sum to be an arbitrary complex number $s$,
\begin{align}\label{eq:zeta_function_reg}
 \energy(s)
 &= \frac{\Omega_{D-2}}{2} \sum_{n=1}^{\infty~}\int_0^{\infty} \frac{dq}{(2\pi)^{D-1}}\,q^{D-2} \left(\frac{q^2}{1+\beta^2q^2}+\frac{\tanh^2\left(\beta\momenta\right)}{\beta^2(1+\beta^2q^2)}+m^2\right)^{s},
\end{align}
and analyzes its meromorphic behaviour in $s$. 
In our case this is useless, given that being the energies finite, whatever power they may have,
they will never be able to compensate for the noncompactness of the space.
Another possibility is to resort to dimensional regularization~\cite{Bollini:1972ui, tHooft:1972tcz}.
The problem is that, in order to obtain a convergent expression for large momentum coordinates,
we would need a negative $D$, generating then an infrared divergence; 
this could be one facet of the well-known UV-IR (ultraviolet-infrared) mixing arising in noncommutative QFTs~\cite{Minwalla:1999px}.

To gain further insight into this comment, 
one can change the coordinates in momentum space, such that the new variables are the momentum eigenvalues given by Eq.~\eqref{eq:realization1};
in this way we obtain
\begin{align}\label{eq:energy_coordinates2}
 \begin{split}
\energy
 &=\frac{\Omega_{D-2}}{2(2\pi)^{D-1}} \sum_{n=1}^{\infty}\int_0^{1/\beta} \frac{dp\,p^{D-2}}{(1-\beta^2p^2)^{D/2+1/2}}\sqrt{\frac{p^2-\beta^{-2}}{\cosh^2\left(\beta\momenta\right) }+\beta^{-2} +m^2}.
\end{split}
\end{align}
After this ``conformal mapping'' it is patent that the problem does not lie in the energies. 
A possibility to employ dimensional regularization would involve introducing an intermediate scale,
which would allow to split the integration interval into two, and afterwards tackle the UV and IR problems separately.
Another idea that arises from this expression, which is the one that we are going to employ later, 
is to simply introduce a cutoff $\Lambda_q$ in the $q$ coordinates, which corresponds to a momentum cutoff
\begin{align}\label{eq:cutoff}
\Lambda:=\frac{\Lambda_q}{\sqrt{1+\beta^2\Lambda_q^2}} .
\end{align}
Although there will still be a divergence coming from the sum, 
it will not depend on the distance between the plates 
and therefore is expected to play no role in physical quantities.
In particular, if we assume that the system will tend to minimize its energy,
according to the principle of virtual work it will experience a pressure 
given by
\cite{Li:2019ohr, Franchino-Vinas:2021lbl}
\begin{align}\label{eq:pressure}
 \pressure:=-\frac{1}{2}\partial_L \energy;
\end{align}
this can readily be  seen to be finite once a cutoff is applied to the integral in the expression of the vacuum energy.

Before dealing with the attributes of the Casimir force in Sec.~\ref{sec:casimir_force}, a comment is in order.
At this point the reader may wonder why there is no explicit measure contribution in Eq.~\eqref{eq:casimir}. 
The fact is that there are some cancellations in the computation, resulting from the normalization of the eigenstates (and the identification of the volume of the plates). 
We can confirm this with the following two arguments. 
The first one, corresponds to the large $L$ limit, 
in which one expects to obtain just the energy density in the whole space, 
i.e. the sum over all the eigenvalues.
This limit can be  computed explicitly, taking into account that in such a limit the series becomes an integral,
\begin{align}\begin{split}\label{eq:energy_coordinates1}
 \frac{\energy}{L}  &=\frac{\Omega_{D-2}}{2} \int_0^{\infty} d{\tilde n} \int_0^{\infty} \frac{dq\, q^{D-2}}{(2\pi)^{D-1}}\, \sqrt{\frac{q^2}{1+\beta^2q^2}+\frac{\tanh^2\left(\beta L\momentatwo{{\tilde n}} \right)}{\beta^2(1+\beta^2q^2)}+m^2}+\mathcal{O}(L^{-1})\\
 &=\frac{ \Omega_{D-2}}{\pi (2\pi)^{D-1}}  \int_0^{\infty}\int_0^{\infty} \frac{dz dq\,q^{D-2}}{\sqrt{1+\beta q^2+\beta^2z^2}}\, \sqrt{\frac{q^2+z^2}{1+\beta^2q^2+\beta^2 z^2}+m^2}+\mathcal{O}(L^{-1}),
\end{split}
\end{align}
where in the last line we have performed a change of variables
\begin{align}\label{eq:change_variables}
z=\beta^{-1}\sinh\left(\beta L\momentatwo{\tilde n} \right) \sqrt{1+\beta^2q^2}.
\end{align}
The leading contribution in Eq.~\eqref{eq:energy_coordinates1} coincides with the result in Ref.~\cite{Mignemi:2017yhd}.
The second argument, is that this expression for the vacuum energy is also obtained in the alternative  Snyder realization~\cite{Franchino-Vinas:2020umq}.


\subsection{The Casimir force}\label{sec:casimir_force}

Notice that the previously derived expression  for the pressure in Eq.~\eqref{eq:pressure} is still not the Casimir force.
The fact is that, as it stands, in general it will be nonvanishing for large $L$, 
which is physically counterintuitive. 
This obstacle can be  remedied by an appropriate subtraction; 
the definition for the Casimir force is thus
\begin{align}
  \pressure^{(C)}: = \pressure(L) - \pressure(\infty),
\end{align}
where the second term in the RHS will generally be an integral of the type that we have 
obtained in Eq.~\eqref{eq:energy_coordinates1}.

In principle one can study the Casimir force in any dimension. 
Here we are going to focus in the $D=1$ case; 
more information on the effect in $D=3$ may be found in Ref.~\cite{Franchino-Vinas:2020umq}.
After the required subtraction, the explicit expression for the Casimir force is 
\begin{align}\label{eq:force^C_D=1}
 \begin{split}
\pressure^{(C)}_{D=1}&=\sum_{n=1}^{\infty} \frac{\momenta}{4 L}\frac{\tanh\left(\beta\momenta\right)}{\cosh^2\argo\sqrt{\tanh^2\left(\beta\momenta\right)+\nada{\beta^2 m^2}}}
-\frac{1}{2\pi \beta^2}\int_{0}^{\infty} dx  \frac{x \tanh\left( x\right)}{\cosh^2 x \sqrt{\tanh^2\left(x \right)+\nada{\beta^2 m^2}}}.  
 \end{split}
\end{align}
Despite the lack of a closed formula, we can show some properties analytically. 
For example, we can show that it is always negative; 
this means that noncommutativity does not change the character of the commutative Casimir force,
regardless of the strength of the $\beta$ parameter.

More insight into the Casimir force can be gained by numerical computations. 
Indeed, thanks to the cosh factors in the denominator, both the series and the improper integral are highly convergent,
so it is not hard to obtain accurate numerical estimations. 
To further simplify  the discussion, consider the dimensionless parameters 
 $\tilde m= \beta m$ and $\tilde L= \beta^{-1}L$. 
 
\begin{figure}
\begin{center}
 \hspace{-1cm}\begin{minipage}{0.49\textwidth}
 \includegraphics[width=1.1\textwidth]{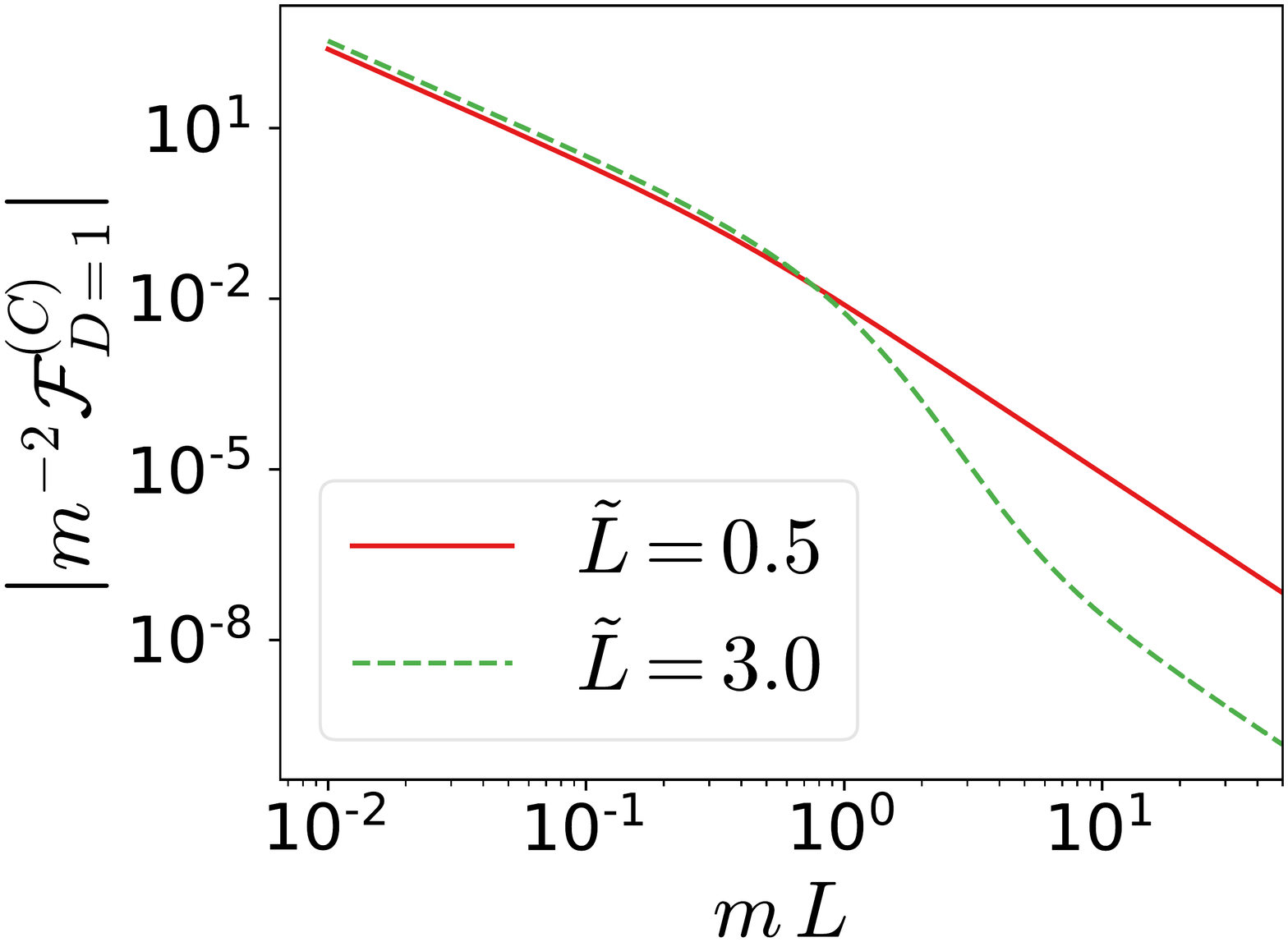}
 \end{minipage}
 \hspace{0.5cm}\begin{minipage}{0.49\textwidth}
 \includegraphics[width=1.1\textwidth]{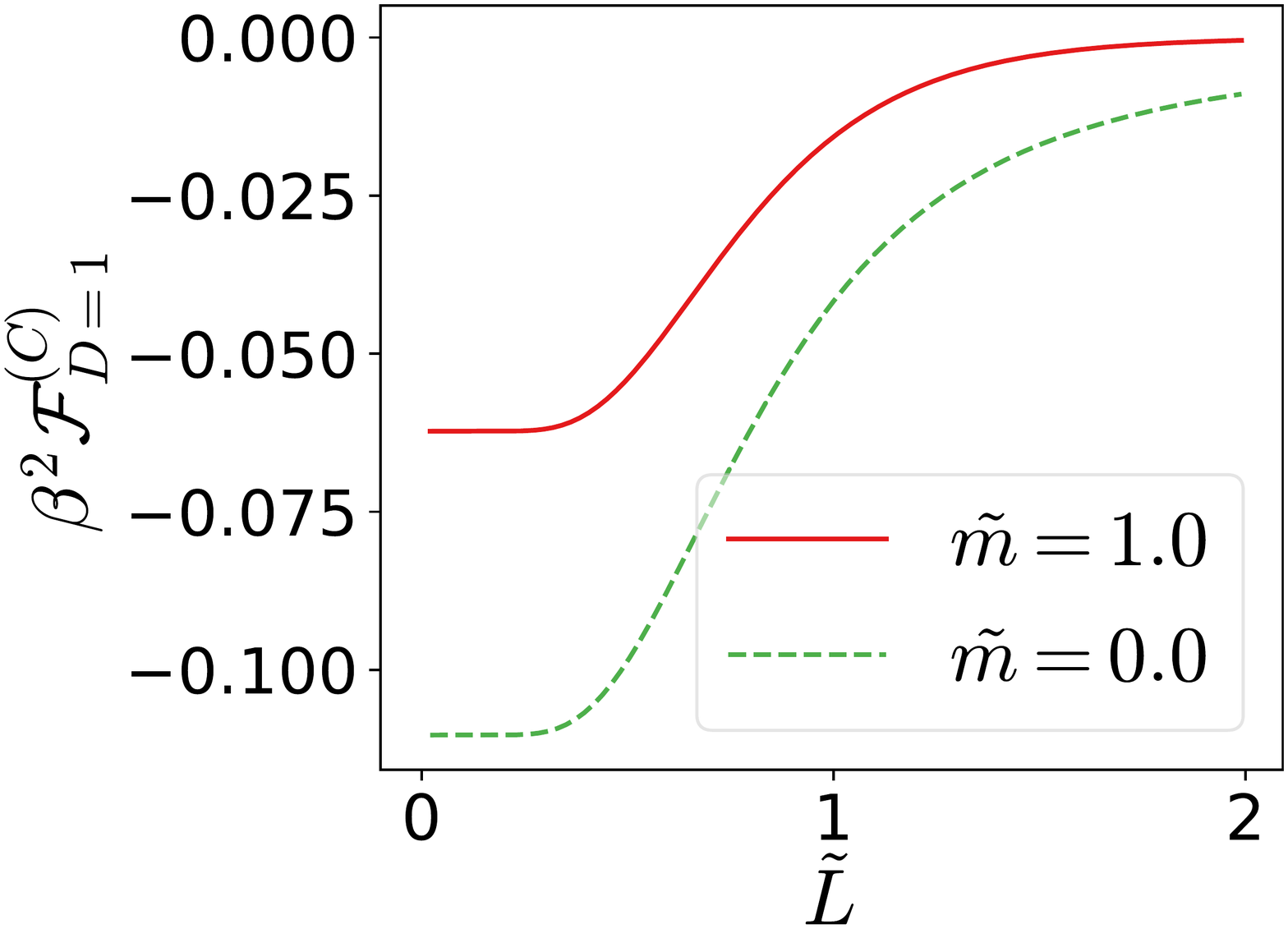}
 \end{minipage}
 \caption{The log-log plot on the left panel corresponds to \nada{$\left\vert m^{-2}\pressure_{D=1}\right\vert$} as a function of $mL$, for $\tilde L=0.5$ (red solid line) and $\tilde L = 3$ (green dashed line). On the right panel, the plot of $\beta^2\pressure_{D=1}$ as a function of $\tilde L$ is shown, for $\tilde m=1$  (red solid line) and   $\tilde m=0$ (green dashed line).}
 \label{fig.forced1}
 \end{center}
\end{figure}

 A peculiar situation arises when we study the pressure in units of mass for fixed $\tilde L$ and as a function of $mL$. 
 In the commutative case, for large $mL$ one obtains an exponential suppression of the force, 
 which is the reason why a preferred attention is usually given to the massless case. 
 In anti-Snyder spacetime, the decay for large $mL$ is only of power-law type (with power minus three).
 The small $mL$ limit is instead not modified, so it behaves as a minus two power.
 Both behaviours can be seen in the left panel of Figure \ref{fig.forced1},
 where we have plotted the pressure for $\tilde L=0.5$ (red solid line) and $\tilde L = 3$ (green dashed line). 
 Notice that this modification in the large-mass behaviour has also been  observed for interacting systems~\cite{Flachi:2020pvn}.

Alternatively, we can study the behaviour of the pressure fixing $\tilde m$ and varying $\tilde L$.
In the infinite $\tilde L$ limit, we obtain a vanishing result, which was expected from the definition of the Casimir force.
On the other hand, in the situation of small $\tilde L$,
there is no divergence as in the naive commutative case; 
instead, we observe that the pressure tends to a constant result. 
This can be understood from the fact that the  parameter $\beta$ should effectively act as an UV cutoff, 
i.e. as a minimum length.
We have depicted this situation in the right panel of Fig.~\ref{fig.forced1}; 
the plot corresponds to the pressure in units of $\beta$,  varying $\tilde L$  for $\tilde m=1$  (red solid line) and   $\tilde m=0$ (green dashed line).

The massless case is analytically more tractable: some factors in the numerator and denominator of the series' terms cancel,
so it is straightforward to use the Euler--MacLaurin formula,
\begin{align}\label{Euler}
\sum_{n=0}^\infty f(n)=\int_0^\infty dn\, f(n)+\left[{\frac{1}{2}}f(n)+\frac{1}{12}{\frac{df(n)}{dn}}-\frac{1}{720}{\frac{d^3f(n)}{dn^3}}+\dots\right]^\infty_0,
\end{align}
and compute the first terms. 
The resulting series is an expansion in inverse powers of $\tilde L$,
what is reasonable if one takes into account that the latter is the only dimensionless parameter and the commutative limit should be obtained for $\beta=0$. 
The pressure, up to the first correction given by the noncommutativity of anti-Snyder space, reads
\begin{align}
\nada{\pressure^{(C)}_{D=1}}\bigg\vert_{m=0}=-\frac{\pi}{96L^2}-\frac{\beta^2\pi^3}{3840L^4}+\mathcal{O}\left(\frac{\beta}{L}\right)^4.
\end{align}


\subsection{Spectral geometry in momentum space?}\label{sec:spectral}
Let us consider for a moment the computation of the Casimir energy in the commutative case. 
Then, the energies to be summed correspond to the eigenvalues of a Laplace type operator $A$,
called the operator of quantum fluctuations, defined over a manifold $M$. 
Using Frullani's formula~\cite{Jeffreys} (or, alternatively, introducing a Schwinger parameter),
one can then relate the vacuum energy with the trace of the heat-kernel $K_A$ of the operator\footnote{One should be careful enough to render valid the formal manipulations in the following equation.} $A$:
\begin{align}\label{eq:vacuum_HK}
 \sum_{n} \omega_n = \sum_{n} \int_0^\infty {\rm d}T \,\frac{e^{-T \omega_n}}{T} = \int_0^\infty  \frac{{\rm d}T}{T} \, K_A(T).
\end{align}
In this expression, the UV divergences correspond to the nonintegrability of $K(T)/T$ at the lower limit of the integral.
The elegance of this approach thus arises from the fact that, under rather general assumptions of regularity of the manifold $M$ and the operator $A$,
the heat-kernel admits an asymptotic expansion for small $T$ in terms of geometric invariants,
\begin{align}
 K_A(T)\sim \sum_{n=0}^{\infty} a_{n} T^{(n-m)/2},
\end{align}
where $m$ is the dimension of $M$.
As an example, the first coefficient ($a_0$) corresponds to the volume of $M$, other to the integral of the Riemann curvature over $M$, 
another to the volume of its boundary (if present), etc.
Replacing this expansion in Eq.~\eqref{eq:vacuum_HK}, 
the final result is that the divergences of the regulated vacuum energy,
in a ``spectral'' dimensional regularization\footnote{By this, we mean shifting the power of $T$ in the denominator of \eqref{eq:vacuum_HK} to have $T^{1+\epsilon}$, being $\epsilon\in\mathbb{C}$ small.}, are given by poles 
whose residues are proportional to the geometric invariants.

Let us now turn to our problem in noncommutative space and ask ourselves: 
does our result for the vacuum energy have any connection with the geometry of the momentum space?
In order to give a partial answer, we first derive expressions for two geometric quantities 
in our momentum space. 
Considering the projective realization \eqref{eq:realization1}, the volume of the hyperbolic space $\mathbb{H}_D$ can be written as
\begin{align}
\text{Vol(}\mathbb{H}_D)= \Omega_{D-1 } \int_0^{\infty} \frac{dq}{\sqrt{1+\beta^2q^2}}\, q^{D-1}.
\end{align}
Additionally, we can change one coordinate, say $q_{\perp}$, into
\begin{align}\label{eq:w}
 \beta w=\text{arcsh}\left(\frac{\beta q_{\perp}}{\sqrt{1+\beta^2q^2_{\parallel}}}\right),
\end{align}
so that the volume of a hyperplane of fixed $w$ is given by the following integral:
\begin{align}
\text{Vol(}\mathbb{H}_{D-1,w=0}):= \Omega_{D-2}\int_0^{\infty} dq\, q^{D-2}.
\end{align}

On the other side, we can recast the expression \eqref{eq:casimir} for the vacuum energy in anti-Snyder space 
in a way such that the divergences are isolated into a few terms,
\begin{align}\label{eq:casimir_NC}
 \begin{split}
\energy
 &= \frac{\Omega_{D-2}}{2 \beta} \sum_{n=1}^{\infty~}\int_0^{\infty} \frac{dq}{(2\pi)^{D-1}}\,q^{D-2} \sqrt{1+\beta^2 m^2- \frac{1}{(1+\beta^2q^2) \cosh^2\left(\beta\momenta\right)}}\\
 &=\frac{\Omega_{D-2}}{2 \beta} \sum_{n=1}^{\infty~}\int_0^{\infty} \frac{dq}{(2\pi)^{D-1}}\,q^{D-2} \sqrt{1+\beta^2 m^2} \left[1 - \frac{1}{2 u}- \frac{1}{8u^2}+\cdots\right],
 \end{split}
 \end{align}
where we have defined $u:=(1+\beta^2 m^2)(1+\beta^2q^2) \cosh^2\left(\beta\momenta\right)$. 
Indeed, increasing the negative powers of $u$ renders the integrals and sums more convergent at infinity, while not affecting the integrability at zero. 
For a fixed dimension $D$, only a finite number of terms in Eq.~\eqref{eq:casimir_NC} will then be divergent.
The first contribution can be worked out in the following way, considering of course an appropriate regularization:
for the series, we will employ the Euler--MacLaurin formula; we will also utilize the change of variables \eqref{eq:w}
and finally obtain\footnote{The boundary contribution appears with the wrong sign in Ref.~\cite{Franchino-Vinas:2020umq}.}
\begin{align}\label{eq:first_coef_hk}
 \frac{\Omega_{D-2}}{2 \beta} \sum_{n=1}^{\infty~}\int_0^{\infty} \frac{dq}{(2\pi)^{D-1}}\,q^{D-2} = \frac{2L}{(2\pi)^{D} \beta }  \text{Vol(}\mathbb{H}_D)  -  \frac{1}{4(2\pi)^{D-1} \beta }  \text{Vol(}\mathbb{H}_{D,w=0}).
\end{align}
Ergo, we see that, if $D<3$, the divergences appearing in our problem are proportional to geometrical quantities in our curved momentum space.
This is remarkably equivalent to the usual considerations in commutative space;
on physical grounds, we expect that one could be able to absorb these divergences in a renormalization 
of the momentum space ``cosmological constant'' and a momentum boundary term of fixed $w$.

We conclude this section by mentioning an important caveat: a geometric description of the higher inverse powers in $u$ is still missing.
Notice also that the large-$u$ expansion in Eq.~\eqref{eq:casimir_NC} is not 
an expansion in a physical parameter; 
this can be seen from Eq.~\eqref{eq:first_coef_hk}, 
insasmuch as it involves one term proportional to $L/\beta$, as well as one proportional to $1/\beta$.


\section{Fermions in Doubly Special Relativity: a geometric way}\label{sec:DSR}
We will now change the subject and discuss another possible application of physics in curved momentum space, 
namely the introduction of fermions in DSR. 
To this end, let us first introduce some essential notions based on Ref.~\cite{Franchino-Vinas:2022fkh}.

First of all, DSR proposes a scenario that may be considered the reverse  of Lorentz Invariance Violation (LIV),
in the sense that a huge difference arises when talking about the fate of Lorentz symmetry:
while in DSR Lorentz symmetry is saved 
at the cost of some deformation~\cite{AmelinoCamelia:2008qg}, 
in LIV one analyzes the consequences of introducing violations to that symmetry in several ways, see~\cite{Mattingly:2005re,Liberati:2013xla,Addazi:2021xuf} and references therein.
Notwithstanding, they share the feature of introducing a minimal length, 
usually believed to correspond to the Planck length $\ell_P \sim 1.6\times 10^{-33}$\,cm, 
as is the case in almost every proposal of quantum gravity;
also, both of them are usually understood as phenomenological theories, 
which do not attempt to describe a full theory of quantum gravity, but rather provide concrete realizations of the first quantum manifestations of spacetime.

The Snyder-type theory that we have studied in the previous section can be cataloged as a DSR theory.
Indeed, the Lorentz invariance is preserved while the algebra of position and momentum operators get deformed in a compatible way.
Correspondingly, one could set the works of Snyder,~\cite{Snyder:1946qz,Snyder:1947nq}, as the initial kick in this area. 
After several years of an almost total oblivion, these ideas regained vigor with the advent of
the $q$-deformed Poincaré algebra~\cite{Lukierski:1991pn},
obtained as a limit of the Drinfeld-Jimbo deformation of the anti-de Sitter algebra $\mathfrak{so}(3,2)$.
Among the subsequent works devoted to the study of particles and fields in $\kappa$-Minkowski and Snyder spaces, we can mention~\cite{Kosinski:2001ii,Dimitrijevic:2003wv,Kowalski-Glikman:2003qjp, Mignemi:2008kn,Govindarajan:2009wt,  Meljanac:2010ps, Girelli:2010wi, Mignemi:2011gr,Meljanac:2017ikx,Mercati:2018hlc, Franchino-Vinas:2018jcs, Arzano:2020jro,Lizzi:2021rlb,Franchino-Vinas:2021bcl}
and especially those dealing with fermions~\cite{Lukierski:1992dt, Giller:1992xg, Nowicki:1992if, Agostini:2002yd, DAndrea:2005hjg}.

In these constructions, one ends up with a deformed dispersion relation and a deformed composition law,
both of them compatible with Lorentz symmetry and 
characterizing the behaviour of particles. 
By deformed composition law, we intend a rule that dictates how momenta corresponding to different particles are to be summed
in order to speak of a ``total momentum'';
for two particles of momenta $q_\mu$ and $p_\mu$, this will be written as $(p\oplus q)_\mu$.
The deformed dispersion relation denotes a relation between energy and momentum of a particle or,
in some sense equivalently, the definition of the mass. 
Of course, the word ``deformed'' in these expressions refers to the fact that the presence of a minimal length 
introduces modifications with respect to the usual laws that govern physics, 
all of which are suppressed by the smallness of such minimal length. 

In the following sections we will try to give a geometric alternative to those constructions, 
taking as a point of departure  the idea of a curved momentum space~\cite{AmelinoCamelia:2011bm,Carmona:2019fwf}
and culminating in the description of  a scalar and a fermionic field. 
Counting the number of independent algebraic quantities that we are trying to describe
(Lorentz invariance and deformed composition law), 
a natural guess is to consider a maximally symmetric momentum space, 
so that a simple correspondence can be established between them and the underlying symmetries of the latter.
Even if several of our formulae will turn out to be rather general, 
we will often stick to a de Sitter space, given its relation to $\kappa$-Minkowski~\cite{Kowalski-Glikman:2002oyi}
and the Snyder algebra~\cite{Snyder:1946qz}.

Before starting, observe the following conventions. We define the Minkowski metric ($\eta_{\mu\nu}$) with mostly minus signs; all other metrics will possess the same signature. Greek indices  are used to label  spacetime components of a tensor ($\mu,\,\nu,\,\cdots=0,1,2,3$), while Latin indices  denote just spatial components ($i,\,j,\,\cdots=1,2,3$). The first Latin characters ($a,\,b,\,\cdots=0,1,2,3$) are employed for components in the local orthonormal frame given by the (inverse) vierbein $e_{\mu}{}^a$. Regarding momenta, i.e. coordinates, we use the following notation: we denote  $p^2:=p_\mu \eta^{\mu\nu}p_\nu$; the set of all the spatial components of a vector $p$ is written as $\vec{p}$ and $\vec{p}^2:= p_i \delta^{ij} p_j$. We use units in which  $\hbar=c= 1$.


\subsection{The modified dispersion relation for a scalar field}\label{sec:scalar}
We shall first study the case of a scalar field.
In SR, the dynamics of the field is governed by the Klein--Gordon  (KG) equation,
\begin{equation}
   \left(\eta^{\mu \nu} \frac{\partial}{\partial x^\nu}\frac{\partial}{\partial x^\mu} +m^2\right) \phi(x)\,=\,0,
   \label{eq:KG_SR}
\end{equation}
which involves the mass $m$ of the particle and the (inverse) Minkowski metric $\eta^{\mu\nu}$.
As is well-known, this equation is Poincaré invariant, 
which is actually the main reason for its derivation;
additionally, it is covariant under the action of diffeomorphisms in spacetime,
given that the term involving derivatives is nothing but the Laplacian.
All its solutions (at least under some regularity condition) can be written as a combination of plane waves, 
i.e. one can work in Fourier space
as long as one remembers to enforce the on-shell condition:
\begin{equation}
   \phi(x)\,=\,\frac{\sqrt{2}}{(2\pi)^{3}}\int  {\rm d}^4 p\, e^{\mathi x^\lambda p_\lambda} 
   \tilde \phi(p)\, \delta(C_{\rm M}(p)-m^2)
   .
   \label{eq:KG_field}
\end{equation}
In this equation we have written the dispersion relation in terms of the Casimir of the Poincaré algebra,
i.e. we have defined
\begin{equation}\label{eq:casimir_M}
  C_{\rm M}(p)\,:=\,p^2\,=\,p_\mu \eta^{\mu\nu}p_\nu.
\end{equation}
This can be interpreted in the following way~\cite{AmelinoCamelia:2011bm}:
the momentum space is flat (encoded in the $\eta$) and $p^2$ is simply the squared distance to the origin.
Then, the Lorentz invariance  in momentum space is realized as the invariance under rotations around the origin,
while the invariance of the KG with the full Poincaré algebra is guaranteed by the additional homogeneity of the momentum space.

With this in mind, the generalization to an arbitrary curved space is immediate.
One can define the Casimir\footnote{The word Casimir is usually employed in the literature as a synonym for dispersion relation.}
to be the squared distance to the origin in curved momentum space, 
i.e. to be one half of  Synge's world function~\cite{Synge:1960ueh} $\sigma(p',p)$ in momentum space (we denote the origin by $p^*$):
 \begin{align}
 C_\text{D}(p)\,:&=\frac{1}{2} \sigma(p^*,p).
  \end{align}
Of course, this definition could be made for any momentum space;
in the case of a maximally symmetric space, however, 
there exists a clear notion of Lorentz invariance for this Casimir.
The action of the Lorentz operator, in general, will depend on the chosen coordinates and will 
not be the one to which we are used to.
In turn, the KG equation may be written as
\begin{equation}
\left(C_{\text{D}}(p)  -m^2\right) \phi(p)\,=\,0.
   \label{eq:KG_DSR_0}
\end{equation}
The invariance properties of the Casimir are inherited by the KG equation
if we assume that the field behaves as a scalar under diffeomorphisms $p\to p^\prime$,
i.e.
 \begin{equation}
\phi^\prime (p^\prime)\,=\,\phi(p).
\end{equation}

One concern is the connection of these formulae to the physics in configuration space. 
In several cases, one can simply resort to the quantum Fourier transform appropriate to the coordinates that have been employed in momentum space~\cite{Mercati:2018hlc} (symmetric, time to the right, time to the left, etc.).
Some of these issues will be addressed in future publications.

Consider now an example of the above-described procedure to fix ideas.
Given the metric of the de Sitter space 
\be
g_{00} (p)\,=\,1,\qquad g_{0i}(p)=g_{i0} (p)\,=\,\frac{p_i}{2 \Lambda}, \qquad g_{ij}(p) \,=\,-\delta^i_j  e^{-p_0/\Lambda}+\frac{p_i p_j}{4 \Lambda^2}, 
\label{eq:metric_alg0}
 \ee
 the distance (and therefore our Casimir) has the expression
 \be
C^{{(S)}}_{\rm D}(p)\,=\, \Lambda^2 \arcch^2 \left(\cosh \left(\frac{p_0}{\Lambda}\right) -\frac{\vec{p}^2}{2 \Lambda^2}\right).
\label{eq:cas_alg_dis}
\ee
We have added the superscript $S$ since, using the formalism of Ref.~\cite{Carmona:2019fwf},
the chosen metric can be linked to the standard basis of $\kappa$-Poincaré~\cite{LUKIERSKI1991331}, also called the symmetric basis.
Indeed, taking into account  the Casimir obtained in the standard basis in the algebraic approach~\cite{Lukierski:1992dt},
\be
  C^{(S)}_\text{A}(p)  \,:=\, \left(2 \Lambda \sinh \left(\frac{p_0}{2\Lambda}\right)\right)^2 ,
\label{eq:kg_alg}
\ee
one can straightforwardly prove that they are related by the equation
\begin{align}
C^{(S)}_{\text{D}}(p)\,=\,\Lambda^2 \arcch^2\left(1 +\frac{C_{\text{A}}(p)}{2 \Lambda^2}\right).
\end{align}


\subsection{The r\^ole of the deformed composition law}
\label{sec:choice_tetrad}
The absence of the deformed composition law in the previous discussions 
will not pass unnoticed to the attentive reader.
After a first cogitation, this sounds not so strange, considering 
that we have only discussed a noninteracting field. 

Following a second reflection, one recalls that the theory of general relativity
describes the interaction between spacetime and any  massive object that lives in it;
from a particle physicist's point of view, 
the interaction takes place as an exchange of momentum with a particle, the graviton.
On physical grounds, we thus expect the situation to be similar in our curved momentum space scenario. 
The scalar field feels the curved momentum space only because it can interact with it;
thus, if the theory should display some consistency, the interaction \emph{should} involve the composition law.
However, in our equations there seems to be no trace of the latter.

The last sentence is a bit naive in view of the fact that, as expressed in Sec.~\ref{sec:DSR},
the composition law is encoded as a symmetry of the metric\footnote{We assume a maximally symmetric space at this point.}.
One consequence of this symmetry is that
\be
g_{\mu\nu}\left(p \oplus q\right)\,=\, \frac{\partial \left(p \oplus q\right)_\mu}{\partial q_\rho} g_{\rho \sigma }\left( q\right) \frac{\partial  \left(p \oplus q\right)_\nu}{\partial q_\sigma},
\label{eq:iso}
\ee
which in the limit $q\to 0$ reduces to
\be
g_{\mu\nu}\left(p\right)\,=\,\left. \frac{\partial \left(p \oplus q\right)_\mu}{\partial q_\rho}\right|_{q\to 0} \eta_{\rho \sigma }  \left. \frac{\partial  \left(p \oplus q\right)_\nu}{\partial q_\sigma}\right|_{q\to 0}.
\label{eq:iso_tetrad}
\ee
Recalling the definition of the vielbein $e^\mu{}_a$, 
 \begin{equation}
g^{\mu\nu}(p) \,=:\,e^\mu{}_a (p)\eta ^{a b}e^\nu{}_b (p),
\label{eq:metric_tetrad}
\end{equation}
it is immediate to realize that the composition law defines one preferred tetrad~\cite{Carmona:2019fwf}
\be
{e}_\mu{}^a (p)\,=\,\delta^a_\nu \left. \frac{\partial \left(p \oplus q\right)_\mu}{\partial q_\nu}\right|_{q\to 0}.
\label{eq:tetrad2}
\ee  

In our previous example employing the symmetric basis of $\kappa$-Poincaré, it is well-known that the deformed composition law is given by 
\be
\left(p \oplus q\right)_0\,=\, p_0+q_0,\qquad \left(p \oplus q\right)_i\,=\, p_i e^{q_0/2\Lambda}+q_i e^{-p_0/2\Lambda},
\label{eq:comp_alg}
\ee
such that the insertion of Eq.~\eqref{eq:comp_alg} into Eq.~\eqref{eq:tetrad2} provides the tetrad associated to $\kappa$-Poincaré in the symmetric basis:
\be
{e}_0{}^0 (p)\,=\,1,\qquad {e}_0{}^i (p)\,=\,0, \qquad {e}_i{}^0 (p)\,=\,\frac{p_i}{2 \Lambda}, \qquad {e}_j{}^i (p)\,=\,\delta^i_j  e^{-p_0/2\Lambda},\quad i,j = 1,2,3.
\label{eq:tetrad_alg}
\ee
 As a consistency check, notice that the metric in Eq.~\eqref{eq:metric_alg0} does indeed correspond to this vierbein.

From the point of view of the dispersion relation, 
the choice of the vierbein and, therefore, 
of the composition law, is irrelevant.
Instead, we will  see shortly that it will turn out to be of fundamental importance in the discussion of the Dirac equation.


\subsection{Dirac equation in curved momentum space}\label{sec:dirac_equation}
In order to construct a Dirac equation, one can follow several guiding principles. 
For example, in Refs.~\cite{Lukierski:1992dt,Donkov:1984fj} the important notion was the fact that it
should be the square root of the KG formula.
Here we will follow the concept that it should be introduced in a geometrical way in momentum space.

The first and probably most crucial obstacle that one encounters 
in such a construction is that one needs a vectorial quantity to be identified with the momentum
in the usual SR.
In fact, in the latter case (and also in curved configuration spaces) the momentum is encoded as a covariant derivative, 
while, in our curved momentum space, the $p_\mu$ are just coordinates.
This can be solved by noting the following~\cite{Synge:1960ueh,DeWitt:1965}:
Synge's world function has the property that its derivatives satisfy
  \begin{align}
f^{\mu}  (p)\,:&=\,\frac{1}{2} \frac{\partial C_\text{D}(p)}{\partial p_\mu},
   \label{eq:f_definition}
   \\
C_\text{D}(p)\,&=\,f^{\mu} g_{\mu \nu }(p) f^{\nu}\,
\label{eq:casimir_metric}.
  \end{align}
Therefore, $f^\mu$ is intuitively the quantity with vectorial character that we were looking for;
taking into account its analogy with momenta in SR, we will call it generalized momentum.

As a next step, we introduce fermions $\psi(p)$ in our theory as quantities 
that transform locally according to the (finite-dimensional) Dirac representation of the Lorentz group, $SO(3,1)$.
This entails, on the one side, bringing into play the vierbein $ e^\mu{}_a(p)$ of the momentum space.
On the other side, the coupling to the generalized momentum is done by the gamma matrices,
which are now defined on our curved momentum space;
following the usual procedure, we therefore define 
the gamma matrices in curved momentum space as
  \begin{equation}\label{eq:gamma_curved}
 \underline{\gamma}^\mu \,:=\,\gamma^a e^\mu{}_a({p}),
\end{equation}
where $\gamma^a$ are the usual gamma matrices in flat spacetime, corresponding to the Dirac spinorial representation.
The definition \eqref{eq:gamma_curved} is such that the anticommutator between gamma matrices gives the identity operator in the internal space times the metric,
  \begin{equation}
\lbrace{ \underline{\gamma}^\mu, \underline{\gamma}^\nu\rbrace}_{}\,=\,2 g^{\mu \nu}(p)\mathbb{I},
\end{equation}
in complete analogy to what happens in curved configuration space~\cite{Birrell:1982ix}. 
This property will be fundamental in proving that the Dirac operator in our definition is also a square root of the KG operator.

Summing up all these ingredients, we write the Dirac equation in curved momentum space as
   \begin{equation}
 \left( \underline{\gamma}^\mu f_\mu({p})-m\right)\psi(p)\,=\,0,
 \label{eq:Dirac_DSR}
\end{equation}
where the generalized momentum with a subscript,  $f_\mu$, is obtained by an appropriate contraction with the inverse metric,
\begin{equation}
 f_\mu(p)\,:=\,g_{\mu\nu}(p)f^\nu(p).
\end{equation}
It is a simple exercise to show that, as previously said,  ``squaring'' the Dirac equation defined in this way 
one obtains the KG equation of the previous section.


\subsubsection{Invariances of the Dirac equation}
One can prove that the Dirac equation in expression \eqref{eq:Dirac_DSR} is covariant under diffeomorphisms in an arbitrary curved space. 
The proof is as follows:
under a combined diffeomorphism $p\to p'$ and Lorentz local transformation $\tilde\Lambda$,  the fermion transforms as
  \begin{align}\label{eq:fermion_local}
     \psi'(p')\,=\,\mathcal{S}(\tilde\Lambda(p))\psi(p).
 \end{align}
 On the other side, the vierbein obeys the rule
 \begin{align}
     e'_{\nu}{}^b(p')=\frac{\partial p'_\nu}{\partial p_\mu} (\tilde \Lambda^{-1})^b{}_a e_{\mu}{}^{a}(p).
 \end{align}
 Moreover, the generalized momentum transforms as a vector, which means that the following equality holds:
 \begin{align}
  e^{\prime\rho}{}_a({p}^\prime)f^\prime_\rho ({p}^\prime)
 &\,=\,
 (\tilde\Lambda^{-1})^{b}{}_{a}(p) e^{\rho}{}_b({p})f_\rho ({p}).
 \end{align}
 Substituting all these transformations in the Dirac equation, 
 we obtain the expression
 \begin{align}
    \Big( \mathcal{S}(\tilde\Lambda(p)) \gamma^a \mathcal{S}^{-1}(\tilde\Lambda(p)) \tilde\Lambda^b{}_a (p) 
    e^{\prime\rho }{}_{b}(p') f^{\prime}_{\rho}(p')-m\Big)\psi^\prime (p^\prime)\,=\,0.
 \end{align}
 One can readily see that covariance is a consequence of the local compatibility condition for the $\gamma$ matrices in momentum space,
 \begin{align}
     \mathcal{S}(\tilde\Lambda(p)) \gamma^a \mathcal{S}^{-1}(\tilde\Lambda(p)) \tilde\Lambda^b{}_a (p)\,=\,\gamma^b.
 \end{align}

For the special case of a maximally symmetric space, 
it can be shown that our  Dirac equation is actually invariant under Lorentz transformations (as defined above).
The interested reader may consult the proof in Ref.~\cite{Franchino-Vinas:2022fkh}.


\subsubsection{An example: the Dirac equation in the symmetric basis of \texorpdfstring{$\kappa$}{k}-Poincaré} \label{sec:dirac_symmetric_basis}

To construct the Dirac equation of $\kappa$-Poincaré we can borrow the results 
that we have previously derived in Sec.\ref{sec:choice_tetrad} for the tetrad, and in Eq.~\eqref{eq:cas_alg_dis} for the Casimir.
Using Eq.~\eqref{eq:Dirac_DSR}, one then immediately obtains the Dirac operator:
\be \label{eq:dirac_symmetric_basis}
\mathcal{D}_{\rm D}^{(S)}\,=\,\frac{ \sqrt{\frac{C_{\text{D}}^{(S)}(p) }{\Lambda ^2}}}
{2\Lambda  \sinh \left(\sqrt{\frac{C_{\text{D}}^{(S)}(p)}{\Lambda ^2}}\right)}
\left[2 \Lambda  e^{-\frac{p_0}{2 \Lambda }} \gamma^i p_i+\gamma^0  \left(2 \Lambda ^2 \sinh \left(\frac{p_0}{\Lambda }\right)-\vec{p}^2\right)\right].
\ee

It is enlightening to compare this result with the findings in Ref.~\cite{Nowicki:1992if}, 
where Hopf algebraic methods were employed. Ref.~\cite{Nowicki:1992if} 
follows the idea in Ref.~\cite{Lukierski:1991pn},
for they consider the standard real form of the  quantum deformation of the anti-de Sitter algebra $\mathfrak{so}_q(3,2)$;
then, the coproducts are deformed in such a way that one of its sectors 
corresponds to a four-dimensional representation of the group $SO_q(3,2)$.
The result they obtain is
\be
\mathcal{D}^{(S)}_\text{Nowicki} \,:=\, \gamma^0 \left(\Lambda \sinh \left(\frac{p_0}{\Lambda}\right)- \frac{\vec{p}^2}{2\Lambda} \right) + e^{-p_0/2\Lambda} p_i \gamma^i.
\label{eq:dirac_alg}
\ee
In order to compare this with our description, notice the following:
in Sec.~\ref{sec:scalar}, we have shown that the Casimir in the Hopf algebraic approach is not the same as the distance in our momentum space. 
What happens if we use the Casimir in Eq.~\eqref{eq:kg_alg} in our derivations? 
Computing the corresponding generalized momentum we simply get
\begin{align}
 \mathcal{D}^{(S)}_\text{A} &=\mathcal{D}^{(S)}_\text{Nowicki},
\end{align}
thus suggesting the equivalence of both approaches.

There are several features that are worth discussing. 
As a first comment, 
``squaring'' $\mathcal{D}^{(S)}_\text{A}$ one does not obtain the associated Casimir $C^{(S)}_\text{A}$; 
instead one has 
\be
\left(
\mathcal{D}^{(S)}_{\text{A}}\right)^2\,=\,C^{(S)}_\text{A}(p)\left(1+\frac{C^{(S)}_\text{A}(p)}{4\Lambda^2}\right).
\ee
This should only enhance the appreciation of our Dirac equation, 
which is always the square root of the KG equation. 
Notice also that there exists a simple link between the Dirac equations obtained with two 
different Casimirs;
as a consequence of the definition of the generalized momentum, 
using the chain rule one is lead to 
\be
\mathcal{D}^{(S)}_\text{A} \,
=\, \underline{\gamma}^\mu g_{\mu \nu}(p) \frac{\partial C^{(S)}_\text{A}}{\partial p_\nu}\,
=\, \mathcal{D}^{(S)}_\text{D} \frac{\partial C^{(S)}_\text{A}}{\partial C^{(S)}_\text{D}}.
\ee

We should also emphasize the r\^ole played by the tetrad. 
It should be clear that, if we had employed a different tetrad, 
we would have obtained a different result,
so that no direct comparison with the algebraic approach could have been made. 
Viewing this from another perspective, 
changing the tetrad (i.e. the composition law) leads to a distinct Dirac equation;
for a particular  example 
comparing $\kappa$-Minkowski and Snyder space, 
we refer the interested reader to Ref.~\cite{Franchino-Vinas:2022fkh}.

\subsection{On the discrete symmetries}\label{sec:discrete}

We conclude the discussion of particles in DSR with a discussion on the discrete symmetries in the proposed scenario.
By discrete symmetries, we mean parity ($\mathcal{P}$) and time reversal ($\mathcal{T}$), which are the discrete Lorentz transformations corresponding to the improper and non-orthochronous sectors of the Lorentz group (recall that this group is not simply connected), as well as charge conjugation ($\mathcal{C}$).
Their proper definition in DSR is a topic that has recently attracted some attention in the literature~\cite{Arzano:2019toz,Carmona:2021pxw}. 
In particular, Refs.~\cite{Arzano:2019toz,Arzano:2020rzu} have discussed the possibility of $\mathcal{PCT}$ violation in the following way: 
if a particle and its antiparticle share the dispersion relation but the momenta of the states are related by the antipode 
of the composition law\footnote{The antipode is simply the inverse as given by the modified composition law; see for example Ref.~\cite{Arzano:2016egk}.},
then one obtains an \emph{a priori} measurable difference in the lifetimes of the particle and its antiparticle. 

To analyze our proposal in curved momentum space,
first consider what happens in flat momentum space. 
Each operator $\mathcal{A}$ acts as a coordinate transformation $L_{\mathcal{A}}$,
as well as on the spin structure, which by abuse of language will be called $\mathcal{A}$, i.e.
   \begin{align}
       \psi_{\mathcal{A}}(p')\,=\, \mathcal{A} \psi (L_{\mathcal{A}} p).
   \end{align}
It is straightforward to show that, in flat momentum space\footnote{We are employing the Dirac representation of the gamma matrices.}, 
the discrete symmetries are realized as
   \begin{align}
        \tilde \psi_{\mathcal{P}}:&\,= \,{\rm i} \gamma^0 \tilde \psi(p_0,-\vec{p}), \label{eq:parity}
    \\
    \tilde \psi_{\mathcal{T}}:&\,= \,{\rm i} \gamma^1\gamma^3 \tilde \psi^*(p_0,-\vec{p}),\label{eq:time_reversal}
       \\
       \tilde \psi_{\mathcal{C}}:&\,=\, {\rm i} \gamma^2 \tilde \psi^*(-p).\label{eq:charge_conjugation}
   \end{align}
   As is the case in flat configuration space, a vital fact turns out to be that the change of coordinates in these operations correspond to isometries of (momentum) Minkowski space. 
   
   Let us then take the Eqs.~\eqref{eq:parity},~\eqref{eq:time_reversal} and~\eqref{eq:charge_conjugation} as definitions 
   and see whether they are satisfied in our curved momentum space scenario. 
   Of course, one might expect that not every curved momentum manifold would admit $\mathcal{PCT}$ as a symmetry,
   which would be analogous to the proven statement that not every curved configuration space possesses it~\cite{Hollands:2002rz, Hollands:2009bke}.
   Considering our Dirac equation, we will say that it satisfies a given symmetry if,
   being $\psi$ one of its solutions, also $\psi_\mathcal{A}$ is. 
   Demanding the existence of parity and time reversal symmetries we are thus lead to the conditions
   \begin{align}\label{eq:condition_discrete_pt}
        e^{\mu}{}_a(p_0,-\vec{p}) f_{\mu}(p_0,-\vec{p})\,=\,
        \begin{cases}
           -e^{\mu}{}_a(p) f_{\mu}(p), \quad &a\,=\,1,2,3,
               \\
               e^{\mu}{}_a(p) f_{\mu}(p), \quad &a\,=\,0.
        \end{cases}
   \end{align}
 On its side, the invariance under charge conjugation, which in our discussion cannot be complete since we have not yet introduced a coupling with an electromagnetic field,  
 enforces the relations
  \begin{align}\label{eq:condition_discrete_c}
       e^{\mu}{}_a(-{p}) f_{\mu}(-{p})\,=\,-e^{\mu}{}_a(p) f_{\mu}(p), \quad a\,=\,0,1,2,3.
  \end{align}

Considering the special case of $\kappa$-Minkowski, 
we can give some arguments to see that $\mathcal{C}$, $\mathcal{P}$ and $\mathcal{T}$ are all 
satisfied under some general assumptions\footnote{Notice that this is a point of disagreement with Ref.~\cite{Andrade:2013oza},
where it was found that $\mathcal{C}$ and $\mathcal{T}$ are not symmetries of their Dirac equation.}
.
An established feature of $\kappa$-Poincaré is that it introduces a fixed time-like vector\footnote{Some people also  consider a generalized version of $\kappa$-Poincaré, introducing a nonnecessarily time-like vector.},
which we will call $n^\mu=(1,0,0,0)$.
Then, if the metric and the tetrad preserve the rotational symmetry in the spatial directions,
we will expect all quantities to depend only on $\vec{p}^2$ and $p_0$ 
(of course, one can trade $\vec{p}^2$ for $p^2$). 
Taking this into account, unless we introduce additional quantities into the theory, 
the contraction of the vierbein and the generalized momentum 
would have the structure
\be
{e^\mu}_a (p) f_\mu (p)\,=\,p_a \bar a_1 \left(\frac{p_\alpha n^\alpha}{\Lambda},\frac{p^2}{\Lambda^2}\right) +  n_a \Lambda \bar a_2 \left(\frac{p_\alpha n^\alpha}{\Lambda},\frac{p^2}{\Lambda^2}\right),
\label{eq:efproduct}
\ee
  where $\bar a_1$ and $\bar a_2$ are in principle arbitrary functions that satisfy some consistency conditions when $\Lambda$ goes to infinity (they should reduce to the SR result). 
  It is immediate clear that this general form do meet the requirements in Eq.~\eqref{eq:condition_discrete_pt}, as well as those in Eq.~\eqref{eq:condition_discrete_c}
  if we perform the additional replacement $\Lambda \to - \Lambda$ when acting with $\mathcal{C}$. 
  
  This additional requirement does  not  seem menacing, since in principle de Sitter space is defined up to the sign of $\Lambda$, 
  so that we are talking of an automorphism in de Sitter space.
  If we think about the  physical meaning of $\Lambda$, i.e. on its r\^ole as an energy cutoff, 
  it is also natural that it should flip its sign when we want to discuss antiparticles, 
  since their energies are reversed in sign.
  Additionally, notice that the introduction of the deformation parameters into the game of symmetry transformations has already been proposed in the past.
 As a few examples, some automorphisms of Hopf algebras have been discussed in Refs.~\cite{Ballesteros:1993zi,Ballesteros_1994},
 while similar transformations have been discussed in QFT~\cite{Kadyshevsky:1983yc, Arzano:2020jro}.


\section{Conclusions and outlook}

With these two examples, the Casimir effect in Snyder space and spinors in DSR, 
we have tried to show that compelling physics and mathematics can still be hidden behind curved momentum spaces.

On the one hand, in the Casimir problem we have seen how an anti-de Sitter metric naturally appears 
when we look for realizations of coordinate and momentum operators in momentum space, 
if these operators ought to be at least symmetric. The problem then basically splits into two main steps
\begin{enumerate}\renewcommand{\theenumi}{(\alph{enumi})}
\item an appropriate way to define boundaries in the noncommutative framework;\label{enum:boundary}
\item the explicit computation of the Casimir force, i.e. of the formal series summing all the available energies.\label{ref:series}
\end{enumerate}
The point \ref{enum:boundary} has been successfully tackled by the implementation of confining potentials.
The issue \ref{ref:series} turned out to be the most interesting, 
since the usual regularization methods (dimensional regularization, zeta regularization, etc.) 
happen to be inappropriate for this problem. 
In fact, in the commutative case there seems to be two different combined divergences: 
the energy modes can be arbitrarily large and the momentum space is noncompact.
In our anti-Snyder case, noncommutativity was able to provide a bound to the energy modes;
however, it has failed to render the momentum space compact. 

In connection to the last point, we have signaled a rather unexpected fact in Sec.~\ref{sec:spectral}: 
the divergences, at least in the lower dimensional cases, 
may be associated with local invariants in the curved momentum space.
This should come with a caveat, 
since working in noncompact spaces in spectral geometry is always more subtle 
than working with compact ones.
Moreover, the interpretation of the different geometrical terms is not clear.
In any case, it will be interesting to see if it is possible to connect these ideas with 
the recent developments of spectral geometry in noncommutative spaces~\cite{Dabrowski:2022ufo,Ponge:2019avw,Iochum:2017ver}.

On the other side,
we have discussed how to properly define fermions in DSR theories. 
Probably the most rigorous way to do so is as in Ref.~\cite{Nowicki:1992if},
i.e. by introducing a finite dimensional representation in one of the coproduct's 
factors of the $SO_q(3,2)$ quantum group.
We have explored an alternative path,
which involves considering a local finite-dimensional representation of the Lorentz group 
in curved momentum space.

\begin{figure}
\begin{center}
 \begin{tikzpicture}[scale=0.60][node distance=1.5cm]

\node (A) at (0, 6) {$\kappa$-Poincaré};
\node (B) at (0, 3) {Hopf algebra};
\node[text width= 4cm] (C) at (0, 0) {\begin{center}
Finite dimensional representation in the coproduct\end{center}};
\node (D) at (8, 0) {Dirac equation};
\node (E) at (8, 6) {dS momentum space};
\node[text width= 4cm] (F) at (8, 3) {\begin{center}
    Local finite dimensional representation
\end{center}};

\draw[->, to path={-| (\tikztotarget)}]
  (A) edge [out=-90, in=90] (B)  ;

  \draw[->, to path={-| (\tikztotarget)}]
  
  (A) edge [out=0, in=180] (E);

\draw[ ->, to path={-| (\tikztotarget)}]
  (B) edge [out=-90, in=90] (C)  ;

  \draw[ ->, to path={-| (\tikztotarget)}]
  (E) edge [out=-90, in=90] (F);

  \draw[ ->,  to path={-| (\tikztotarget)}]
  (C) edge  [out=0, in=180] (D)  ;

  \draw[ ->,  to path={-| (\tikztotarget)}]
   (F) edge [out=-90, in=90] (D);

  \end{tikzpicture}
  \end{center}
\caption{A diagram for the conjectured equivalence between the definitions of Dirac spinors in $\kappa$-Poincaré.}
\label{fig:diagram_k}
\end{figure}
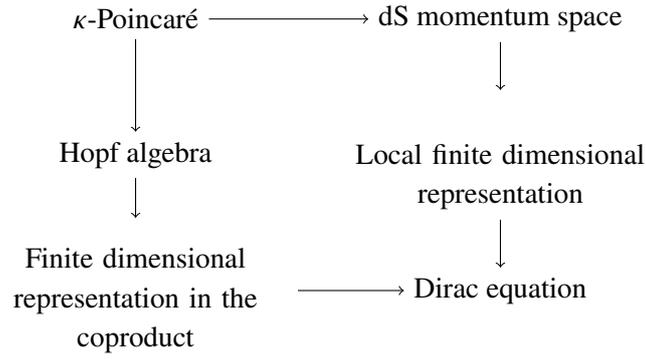

In this framework, we have studied the equivalence of both approaches
in the case of the symmetric basis in $\kappa$-Poincaré,
so that, surprisingly,  the diagram shown in Fig.~\ref{fig:diagram_k} seems to commute. 
This leads to the question: does this diagram commute in other examples? How general is this relation? 

One further point that deserves comment is the r\^ole of covariance in our discussion. 
In the Casimir case, we have proved that the same result is obtained working with two 
different realizations (or equivalently coordinates) in momentum space.
For the study of fermions, all our results turn out to be covariant in momentum space.
This is contrary to some discussions frequently found in the literature, 
where physical quantities are claimed to depend on the choice of realizations/coordinates;
in our perspective, 
this may be due either to the fact that the chosen quantity is actually  not physical, or to erroneous manipulations.


Of course further work is required, probably in all the directions discussed in this article. 
In the meantime, we expect this article to motivate the reader to 
delve into the mathematical notions that we have presented, 
or, at least, to appreciate the beauty of curved momentum space.


 \acknowledgments
 SAF is grateful to A.~P.~Balachandran, J.~Lukierski and A. Sitarz for their questions and comments.
 SAF acknowledges the support of Helmholtz-Zentrum Dresden-Rossendorf, Project 11/X748 (UNLP)
 and 
 PIP 11220200101426CO (CONICET).   
 JJR acknowledges support from the
Unión Europea-NextGenerationEU (``Ayudas Margarita Salas para la formación de jóvenes doctores''). This work has been partially supported by the Agencia Estatal de Investigaci\'on (Spain)  under grant  PID2019-106802GB-I00/AEI/10.13039/501100011033, by the Regional Government of Castilla y Le\'on (Junta de Castilla y Le\'on, Spain) and by the Spanish Ministry of Science and Innovation MICIN and the European Union NextGenerationEU/PRTR.
The authors would like to acknowledge the contribution of the COST Action CA18108 ``Quantum gravity phenomenology in the multi-messenger approach''.

 \appendix 
 
\bibliographystyle{JHEP}
\bibliography{biblio_dsr,biblio_casimir}

\end{document}